\documentclass[a4paper,noarxiv,onecolumn]{elsarticle}

\usepackage[utf8]{inputenc}
\usepackage[english]{babel}
\usepackage[T1]{fontenc}
\usepackage[ruled,vlined]{algorithm2e}

\usepackage{graphicx}
\usepackage{amsmath}
\usepackage{bbold}
\usepackage{xcolor}

\usepackage{hyperref}
\usepackage{breakurl}

\journal{Computer Physics Communications}

\begin{document}

\begin{frontmatter}



\title{Highly optimized quantum circuits synthesized via data-flow engines} 


\author[a,b]{P\'eter Rakyta}
\affiliation[a]{organization={Department of Physics of Complex Systems, E\"otv\"os  Lor\'and  University},
            addressline={Pázmány Péter sétány 1/a}, 
            city={Budapest},
            postcode={1117}, 
            state={},
            country={Hungary}}
\affiliation[b]{organization={Wigner Research Center for Physics},
            addressline={29–33 Konkoly–Thege Miklos Str}, 
            city={Budapest},
            postcode={1121}, 
            state={},
            country={Hungary}}

\author[c]{Gregory Morse}
\affiliation[c]{organization={Department of Programming Languages and Compilers, E\"otv\"os  Lor\'and  University},
            addressline={Pázmány Péter sétány 1/a}, 
            city={Budapest},
            postcode={1117}, 
            state={},
            country={Hungary}}

\author[a]{Jakab N\'adori}

\author[b]{Zita Majnay-Tak\'acs}

\author[d]{Oskar Mencer}
\affiliation[d]{organization={Maxeler Technologies, a Groq company},
            addressline={16192 Coastal Hwy}, 
            city={Lewes},
            postcode={19958}, 
            state={Delaware},
            country={United States}}

\author[b,e]{Zolt\'an Zimbor\'as}
\affiliation[e]{organization={Algorithmiq Ltd},
            addressline={Kanavakatu 3C}, 
            city={Helsinki},
            postcode={00160}, 
            state={},
            country={Finland}}          

\begin{abstract}
The formulation of quantum programs in terms of the fewest number of gate operations is crucial to retrieve meaningful results from the noisy quantum processors accessible these days.
 In this work, we demonstrate a use-case for Field Programmable Gate Array (FPGA) based data-flow engines (DFEs) to scale up variational quantum compilers to synthesize circuits up to $9$-qubit programs.
 This gate decomposer utilizes a newly developed DFE quantum computer simulator that is designed to simulate arbitrary quantum circuit consisting of single qubit rotations and controlled two-qubit gates on FPGA chips.
 In our benchmark with the QISKIT package, the depth of the circuits produced by the SQUANDER package (with the DFE accelerator support) were less by $97\%$ on average, while the fidelity of the circuits was still close to unity up to an error of $\sim10^{-4}$.
\end{abstract}


\begin{highlights}
\item An FPGA based accelerator is developed and utilized in quantum gate decomposition experiments. We demonstrated significant quantum circuit compression using the adaptive compression algorithm in conjunction with the developed high-performance unitary simulator.
\end{highlights}


\end{frontmatter}

\section{Introduction}

The quest to demonstrate quantum advantage in solving useful computational problems is a driving force motivating the researchers to engineer better hardware providing more reliable gate operations, efficient methods of controlling the hardware resources and approaches to compile quantum programs utilizing the fewest logical operations.
All of these efforts are spent for the sake of scaling up quantum computation to exhibit quantum advantage in solving computational problems on the current noisy intermediate-scale quantum (NISQ) devices.
Some algorithm classes are already proven to provide significant speedup compared to the best classical algorithms. 
The proposed applications are primarily based on the most well-known quantum algorithms such as Shor's  period finding routine \cite{doi:10.1137/S0097539795293172}, Grover's search \cite{PhysRevLett.79.325}, and
the HHL method for  solving linear systems of equations \cite{PhysRevLett.103.150502,doi:10.1137/16M1087072,PhysRevLett.122.060504}.
In addition, a different class of algorithms tailored by the combination of quantum and classical resources are believed to show practical advantage over classical computing as well.
In recent studies, promising schemes were proposed to solve computational problems based on variational quantum circuits, like the quantum approximate optimization\cite{2014arXiv1411.4028F,harrigan2021quantum}, the quantum eigensolver \cite{Peruzzo2014,Kandala2017,arute2020hartree}, and the variational solver for linear equations \cite{XU20212181} and differential equations \cite{kyriienko2021solving,Leong2022}.

Naturally, in calculations carried out on NISQ devices, the error of the circuits is growing with the number of gates \cite{harrigan2021quantum}. This pinpoints the crucial role of quantum circuit optimizations.
In this respect one should note that most gate based NISQ processors exhibit limited connectivity between the qubits; this provides additional constraints for quantum gate synthesis.
Depending on the underlying architecture\cite{Linke3305,Proctor2022,Hu2022} several techniques were developed to adjust quantum circuits according to the qubit connectivity layout \cite{9349092,Saki2023,2020arXiv200304462Y,rakyta2022approaching,9743148,https://doi.org/10.48550/arxiv.2206.07885,https://doi.org/10.48550/arxiv.2206.13645}, making them executable on NISQ processors.
While qubit swaps are quite expensive in terms of additional gate operations, optimization based quantum gate synthesis tools were shown to keep the gate count low in such problems \cite{9259942,smith2021leap,rakyta2022approaching,squander2}.
However, the optimization process is computationally quite demanding and no use cases were demonstrated involving more than $6$ qubits, unless limited to shallow circuits \cite{https://doi.org/10.48550/arxiv.2205.04025,https://doi.org/10.48550/arxiv.2210.09191,https://doi.org/10.48550/arxiv.2301.08609}.
In nutshell, quantum gate synthesis for NISQ devices can be summarized in two main points:
(i) the routing of the single- and two-qubit gate operations poses a discrete combinatorial problem \cite{Sivarajah_2020,davis2019heuristics,9259942,patel2021robust,9651462,4082128,smith2021leap,https://doi.org/10.48550/arxiv.2111.11387,10.1145/3519939.3523433}, 
(ii) while finding the appropriate rotational angles for the single-qubit gates implies an optimization problem over continuous variables.
Previous studies have designed various numerical strategies to resolve these computational tasks, either by bypassing the combinatorial problem by static gate structures \cite{madden2021best,rakyta2022approaching}, or by turning it into continuous variable optimization problem \cite{2020arXiv200304462Y,squander2,https://doi.org/10.48550/arxiv.2205.01121}.

In this work, we present a novel approach enabling us to scale up the gate synthesis approach of Refs.~\cite{squander2,https://doi.org/10.48550/arxiv.2205.01121} in the number of the involved qubits, while still obtaining highly optimized circuits with fidelity close to unity.
{The primary contributions of our current work to the field of quantum compilation can be summarized through the following key points:
\begin{itemize}
    \item Due to exponentially scaling computational complexity, optimization-based quantum compilers are hardly being applied on problems involving larger count of qubits. 
    In this work, we developed a hardware accelerator to increase the computational performance of optimization-based quantum compilers.
    \item We have designed a data-flow engine (DFE) formulation of single-qubit and controlled two-qubit quantum gate operations.
    \item We developed a universal quantum computer simulator on FPGA chips, implementing a DFE design for a sequence of gate operations.
    \item We incorporated the developed DFE quantum computer simulator into the SQUANDER package and showed that FPGA accelerators can be used to solve practical problems in the field of quantum gate synthesis with efficiency not reported previously.
\end{itemize}}
Our numerical experiments on $6{-}9$ qubit unitaries -- taken from the online database \cite{ibm_mapping} containing series of circuits published as part of the Qiskit Developer Challenge, a public competition to design a better routing algorithm -- resulted in $7$-$270$ times less $CNOT$ gate count than obtained by the \emph{transpile} function of the QISKIT package \cite{qiskit_org} with optimization level set to $3$.
In addition, we realized significant compression relative to the circuits present in the online database.  
(Our results are summarized in Table \ref{table:results}. The quantum circuits reported here are provided within the SQUANDER package in QASM format accessible via a GitHub repository\cite{SQUANDER_github}.)

Our numerical approach relies on the peculiar computational capabilities of the Field Programmable Gate Arrays (FPGAs).
The so-called data-flow programming model\cite{10.1145/1013208.1013209}
enables one to realize computational concurrency that is fundamentally different than traditional parallel programming on GPU and CPU hardware.
While on GPU/CPU architectures it is a good programming strategy to split the workload into concurrent computing tasks (like the computation of the gradient components of the cost function, or launching the optimization process in concurrent instances with multiple initial parameter sets \cite{davis2019heuristics,9259942,smith2021leap,https://doi.org/10.48550/arxiv.2205.01121}), the data-flow programming model can be utilized to introduce parallelism into computation tasks with high degree of data dependency.
The basic concept of the data-flow programming can be explained by thinking about a stream of computational data flowing through a hardware resource performing a fixed logical transformation on the elements of the stream, transforming a single data element in each clock cycle.
By chaining up such hardware resources we realize a data-flow computing model: while a hardware element is transforming the $i$-th element in the stream, the next hardware element in the chain is already working on the previously transformed, $i{-}1$-th element of the stream. 
By providing long enough data stream to the hardware one can realize an efficient parallel execution, even if the underlying computing model exhibit data-dependency.
As we can see the operation of quantum gates provides a good example to be simulated via such computing model.
Namely, when it comes to the emulation of quantum computers, the quantum gate operations need to be applied on the emulated qubit register one after another in a chain to ensure the correctness of the calculations, unless some of the gates are combined into composite operations.
Even if the gates commute with each other, computationally it is not safe to apply them in parallel on the quantum state stored in the memory of a computer, since concurrent transformations might interfere with each other. (In general, the gate operations are mixing memory-distant amplitudes, hindering the correctness of the implementation when the concurrent transformations overlap.)
For such calculations the FPGA based DFEs provide a natural choice to speed up the calculations.
The idea to emulate quantum computers on FPGAs is not a novelty, the foundations of such implementations were laid down by the first proposals for circuit and gate implementation almost two decades ago.
The first demonstrations implementing a whole circuit on the FPGA fabric were performed only for a few qubits \cite{1347938} and for a limited number of gate operations \cite{4556828}.
A different approach was proposed by Ref.~\cite{7250404} providing a single on-chip instance of each implemented gate and streaming the state vector through these components, while Refs.~\cite{Pilch2019,Goto_2007} combined several elementary gate operations in the circuit into a single composite matrix operation.
Specific implementations designed to emulate quantum algorithms like the quantum Fourier transform and Grover's search \cite{Lee2016,Fujishima_2003,Mahmud2020}, quantum Haar transform \cite{Mahmud2019}, quantum k-means classification algorithm \cite{Bonny2020}, computational fluid dynamics \cite{10.3389/fmech.2022.925637}, or quantum machine learning \cite{https://doi.org/10.48550/arxiv.2206.09593} were also demonstrated by several research works during the years.
However, none of these implementations provide enough variability to simulate quantum circuits arising in our gate synthesis approach without reprogramming the FPGA chip or expensive CPU pre-processing.
In order to fulfill this gap we developed our own solution for the simulation of universal quantum computers.

{In general, FPGA development work entails significant involvement in low-level hardware control tasks, demanding notable expertise in this domain. It is a rare occurrence that such low-level expertise can directly contribute to meeting high-level demands, such as quantum gate compilation. An essential message conveyed here is the capacity of higher-level development frameworks like MaxCompiler to serve as a bridge between different domains of research, highlighting promising applications of FPGA hardware in quantum computing research.}

\section{Description of the DFE implementation}

By optimization based quantum gate synthesis a quantum program $U$ is approximated by a quantum circuit described by unitary $V$ and parameterized by the rotational angles of the qubits. 
To quantify the distance $U$ and its approximation $V$ we used a metric based on the Frobenius norm.:
\begin{equation}
    f(U,V) = \frac{1}{2}\left\|V-U \right\|_F^2 = d - \textrm{Re}\left[{\textrm{Tr}}(VU^{\dagger})\right], \label{eq:frobenius}
\end{equation}
with $d$ being the dimension of the unitaries.
(For more details see the appendix.)
Metric (\ref{eq:frobenius}) can be used as a cost function in the optimization problem to approximate $U$ by $V$, since $f(U,V)\rightarrow0$ when $V$ is close to $U$ and $f(U,V)>0$ otherwise.
The most expensive task in the evaluation of (\ref{eq:frobenius}) is to obtain the matrix $V$ itself.
The computational complexity to calculate an $n$-qubit matrix $V$ via $M$ gate operations is $\mathcal{O}(M\cdot 4^n)$, since in general each element of the $2^n\times 2^n$ matrix $V$ is involved in the gate operation.
If there are $P$ free parameters associated with the rotation angles of the gates in the circuit, the complexity of making a single update on the parameter set in the ADAM algorithm \cite{https://doi.org/10.48550/arxiv.1412.6980} is $\mathcal{O}((P+1)\cdot M\cdot 4^n)$, since $P$ gradient components need to be calculated. In addition, the value of the cost function is needed to test whether to terminate the optimization process or not.

\paragraph*{Basic concept of a gate operation via DFE:}

Since the execution time scales exponentially with the number of qubits involved in the quantum program, optimization based quantum compilers were so far demonstrated only on smaller sized circuits up to $6$ qubits \cite{davis2019heuristics,9259942,https://doi.org/10.48550/arxiv.2205.01121,smith2021leap,2020arXiv200304462Y,madden2021best}.
\begin{figure}
     \centering
     \includegraphics[width=0.9\textwidth]{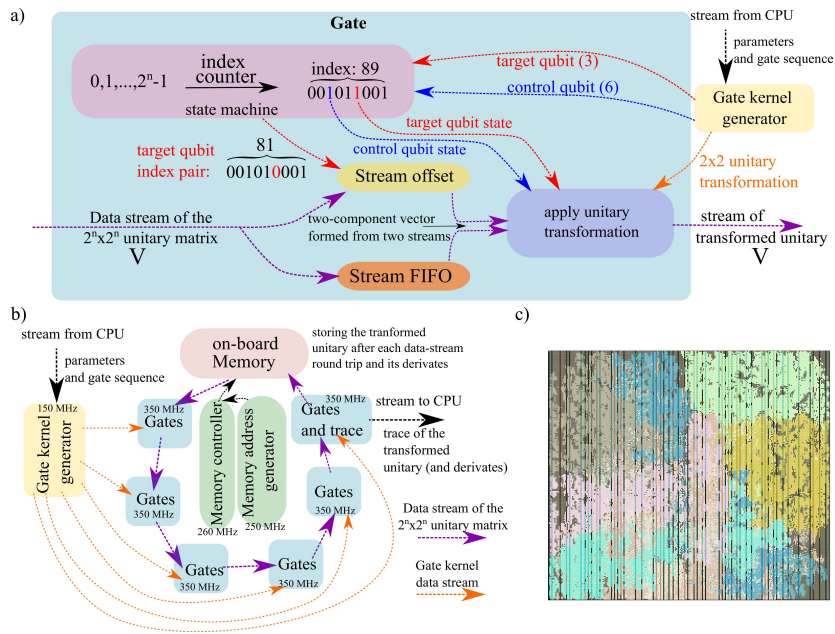} 
     \caption{a) The design of a \emph{Gate} block performing universal single qubit rotations and controlled two-qubit operations. 
     The data stream $V$ is split and offset to combine the elements in the columns of the unitary according to the current gate transformation. 
     b) The implementations of the \emph{Gate} blocks are chained up into a data-flow cycle, each gate performing a single gate operation (or it's differentiation with respect to one of it's free parameters) on the data stream $V$. The transformed unitary is streamed into the on-board memory unit. During the very last gate operation the trace of the unitary is calculated and streamed back to the host CPU.
     c) The data-flow implementation of the design on one of the SLR blocks, indicating the different implementation blocks with coloring. The plotted SLR region contains $108$ gate implementations that can be executed in parallel along a single unitary data-stream. The FPGA chip of the Xilinx Alveo U250 card contains $4$ of these SLR region increasing the concurrency of a single FPGA chip up to $432$ concurrent gate operations.}
     \label{fig:building_blocks}
 \end{figure}
In order to speed up the evaluation of the cost function and extend the scope of the optimization based gate synthesis approach to larger circuits we developed a data-flow implementation for FPGA chips to perform general single- and controlled two-qubit gate operations. 

\begin{figure}
     \centering
     \includegraphics[width=0.9\textwidth]{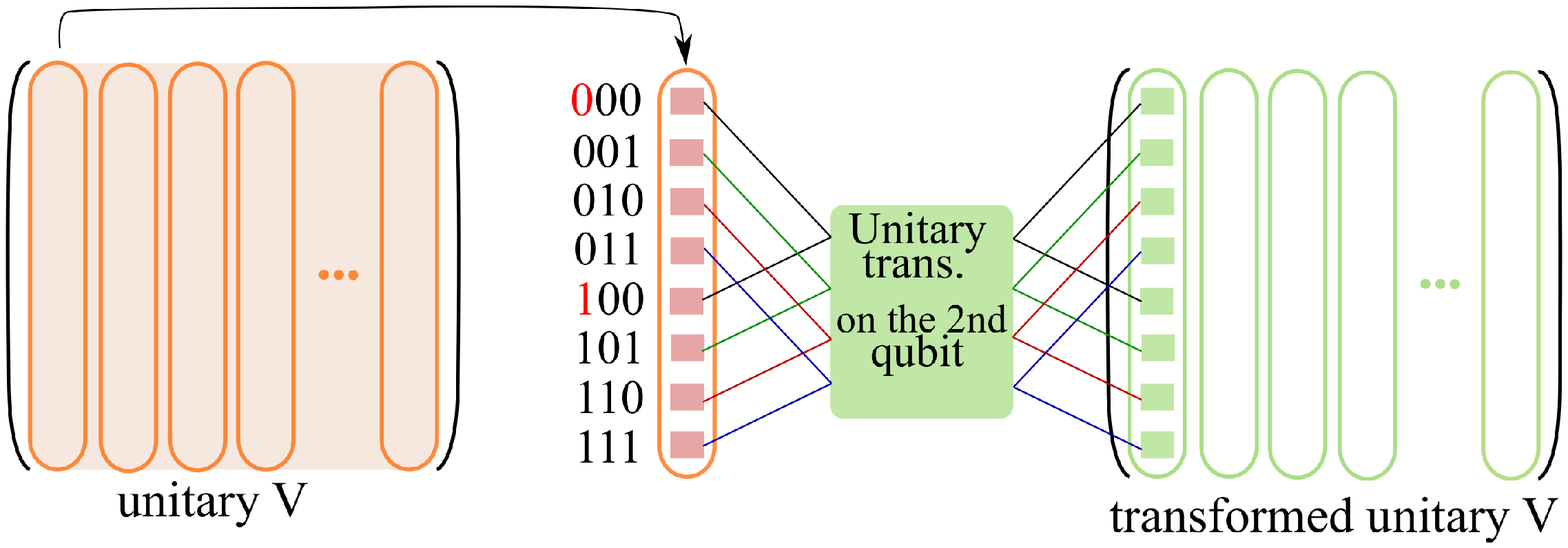} 
     \caption{The gate transformations are applied column-wise on the unitary $V$. The elementary gate operations can be represented by sparse unitary transformations. In the example shown in the figure, a single qubit rotation is performed on the third qubit. Assuming the unitary $V$ given in computational basis, the unitary transformation combines those elements in the column, whose indices differ only in the most significant bit. (The state of the remaining qubits are identical.) }
     \label{fig:column_operation}
 \end{figure}

Figure \ref{fig:building_blocks}.a) shows the scheme of the universal gate implementation.
The unitary (or several columns of it) is streamed through the gate implementation element-wise in column-major order. 
(As shown in Fig. \ref{fig:column_operation}, the gate operations on the unitary $V$ are applied column-wise via sparse unitary operations.)
The unitary operation associated with a quantum gate is applied on this data stream.
To this end a gate kernel generator block elaborates the data up-streamed from the CPU describing the quantum circuit. 
The gate kernel generator block streams the computational data, such as the $2\times2$ unitary ("gate kernel") that should be applied on the target qubit and the labels of the target and control qubits associated with the operation, to the block implementing the actual gate operation. 
(The \emph{Gate} block holds these data during the gate operation. 
{As we will discuss at later, MaxCompiler provides us with high level interfaces mapped to hardware elements, including a stream-hold component for this specific purpose.)}
The $2\times2$ gate kernel can be given by a general single-qubit rotation\cite{qiskit_org}
\begin{equation}
   U^{2\times2}  = \begin{pmatrix}
         {\rm cos}\left(\theta/2\right) & -e^{{\rm i}\lambda}{\rm sin}\left(\theta/2\right) \\ e^{{\rm i}\phi}{\rm sin}\left(\theta/2\right) & e^{{\rm i}(\lambda+\phi)}{\rm cos}\left(\theta/2\right)
   \end{pmatrix},
\end{equation}
where $\theta$, $\phi$ and $\lambda$ are the free parameters of the rotation.
In order to apply the transformation, one need to identify and combine the pairs of elements in the stream of $V$ corresponding to basis vectors that differ only in the target qubit state (see Fig.~\ref{fig:column_operation}. for an example).
This is achieved by splitting the unitary stream with a multiplexer and offsetting one of the streams by a dynamic offset generated via an index counter state machine, so the two streams would provide the appropriate pairs for the transformation.
(See the purple box in Fig. \ref{fig:building_blocks}.a) labeled by "apply unitary transformation".)
A state machine is a block on the FPGA chip that determines its state according to the state in the previous clock cycle by iterating the implemented logical circuit.
The definition of the state machine implies that all the involved logical operations are done in a single clock cycle {(without inserting pipeline registers into the circuitry)}, hence timing constraints poses limits on the maximal complexity of these components. 
{MaxCompiler provides dedicated components to design state machines and integrate them into computing kernels.}
In our particular case the main role of the index counter state machine is to iterate over the basis vectors spanning the Hilbert space of the $n$-qubit register, i.e. $I=0,1,\dots2^n-1$, and to determine the dynamic offset discussed above.
The offset parameter is determined from the binary representation of $I$ by negating the bit corresponding to the target qubit.
Labeling the state $\left|\downarrow\rangle\right.$ by $0$ and $\left|\uparrow\rangle\right.$ by $1$ the binary representation of $I$ can be directly associated with the basis vectors of the multi-qubit system.
Figure \ref{fig:building_blocks}.a) shows a specific example in which the target qubit of the gate operation equals to $3$ in a $n=9$ qubit register. 
Since the labeling of the (qu)bits starts with $0$, the target qubit can be associated with the $4$-th bit from the right in the binary representation of the counter index (see the red colored bit in the pink area of Fig. \ref{fig:building_blocks}.a)).
By negating the bit of the target qubit we obtain the basis vector differing from $I$ only in the state of the target qubit.
The actual offset parameter to be applied on the stream is the difference between these two index values, being $-2^t$ with $t$ labeling the target qubit. 
(In the example shown in Fig. \ref{fig:building_blocks}.a) $t=3$ and the offset parameter is $-2^3 = 81-89$.)
In general, the offset parameter is $2^t$ if the target qubit state is labeled by $0$ in $I$, and $-2^t$ if it is $1$.
When working with an $n$-qubit register, the largest offset parameter would be $2^{n-1}$ if a quantum gate is applied on qubit $t=8$.
{The secondary role of the state machine is to select either the upper or lower row of the gate kernel $U^{2\times 2}$ to be used in the transformation of the amplitudes.
This decision is based on the state of the target qubit corresponding to the amplitude index in question.}
The implementation of a dynamic offset is memory intensive, designs supporting larger offsets consume a significant portion of the on-chip memory.
In our implementation the dynamic stream offset needs to store $2^n$ values to support the maximal, $\pm2^{n-1}$ offset to cover all possible single gate operations on an $n$-qubit register.
This constraint is one of the decisive factors when it comes to scale up the implementation, since the amount of the accessible on-chip memory is limited at several megabytes at most.
While the data stream of $V$ is designed to carry a single matrix element in each tick, the stream offset requires $2^{n-1}$ clock cycles to be filled with data elements at the beginning of the execution.
In order to balance the latency of the stream offset, the other arm of the unitary stream is delayed by an $2^{n-1}$ deep FIFO (first-in-first-out), so the two data streams arrives in the same clock cycle to the block performing the unitary transformation (see the purple box in Fig. \ref{fig:building_blocks}.a)).
Here we notice, that the applied FIFO blocks requires significant amount of on-chip memory as well.
{Stream offsets offer the capability to intentionally use a signal after a specified delayed number of clock cycles, either at the location of the offset signal itself (utilizing a negative offset) or on all paths except the offset signal (employing a positive offset). This functionality enables the retrieval of past or future values from the input stream by postponing data until the most recent requested value becomes accessible. Maxcompiler provides the means to apply OFFSET objects and FIFOs to data streams, facilitating the implementation of this feature.
When dealing with dynamic offsets, where the offset value may change during program execution, the most efficient method to implement this functionality is by utilizing on-chip memory.
The Xilinx Alveo U250 FPGA card used in our work offers two types of on-chip memories. The dual-port BlockRAM modules can store up to $36$ Kbits of data (composed of two $512$ deep and $18$ bit wide BRAM18 components) and can be used as one or two independent memory units. In both cases they consist of two read and two write ports and it is possible to tile them into different depth and width configurations. The UltraRAMs (URAMs), on the other hand, can store up to $288$ Kbits of data, but has only one single write and one read port. Additionally, URAMs can be used only as $72$ bit wide and $4096$ deep memories and some specific functions, e.g., dual clock FIFO implementations, are not supported on these elements.
(Therefore, URAMs can not be used in FIFO buffers joining the computing kernels driven by different clock frequencies.) In total, Alveo U250 provides $2000$ BRAMs and $1280$ URAMs adding up to $54$ MBs of on-chip memory.  
The utilization of the available memory is the primary limiting factor in scaling up our design.}
Finally, the elements of the transformed unitary leaves the \emph{Gate} block in a single output data stream, indicated with the purple stream on the right of Fig. \ref{fig:building_blocks}.a).

The logic in the block of the unitary transformation is governed by the data streams encoding the state of the target and control qubits, and the data stream carrying the elements of $U^{2\times2}$ describing the transformation on the target qubit. 
(All the four elements of $U^{2\times2}$ arrive to the transformation block in the same clock cycle over four parallel stream channels.)
According to the state of the target qubit either the first or the second row of $U^{2\times2}$ is applied in the transformation of the stream $V$.
The state of the control qubit is used to simulate controlled two-qubit transformations: in case the control qubit is in state $\left|\downarrow\rangle\right.$ (bit $0$) the corresponding element of $V$ is leaved unchanged. 
Otherwise the elements are transformed according to $U^{2\times2}$.
We also notice that the outlined design can be used to evaluate the derivative of a gate operation as well. 
In this case the elements of $U^{2\times2}$ are set to their partial derivatives with respect to one of the free parameters.

\paragraph*{Scaling up computational concurrency to simulate quantum circuits:} 

To simulate quantum circuits on DFEs we chain up multiple instances of the outlined \emph{Gate} blocks as indicated in Fig. \ref{fig:building_blocks}.b).
Along the chain the output stream $V_{\rm out}$ of the \emph{Gate} blocks are connected to the input stream $V_{\rm in}$ of the next \emph{Gate} block, while the $2\times2$ gate kernel matrices are streamed from a single \emph{gate kernel generator} block.
{Each of the \emph{Gate} blocks are implemented using the kernel block abstraction of the MaxCompiler. 
Kernels are graphs of pipelined arithmetic units expressing functionality on the
level of mathematical operations, on flow control levels, and on stream IO operations. This description is mapped to a hardware implementation on a DFE by MaxCompiler.
The flow control, such as the distribution of clock enable signal, the filling and flushing the kernel when the computation is starting or ending, the stalling mechanism when no new data enters or leaves the kernel, is automatically handled by the development framework, abstracted away from the user. 
As a result, the input and output stall latency is matched between kernel streams taking into account the pipeline depth of the kernel and the stalling logic, evaluated internally with MaxCompiler. 
(In complex designs where many kernels are connected to each other the stall mechanism needs to be orchestrated not to loose valid data during the execution and to not produce false data on the output. For example, a kernel should stop sending data to the input of another kernel if the receiver has got into a stalled mode. 
Since the stall signal needs to propagate up (or down) over a stall logic, each input and output has its own stall latency, describing how many clock cycles shall be able to keep elaborating data, after the stall signal has been asserted on it. MaxCompiler sets up the necessary machinery automatically.)
In our implementation, we manually ensure pipelining stages after at most $2$ arithmetic or logic operations.
Additionally, instead of relying on a single large fanout, we organized the clock enable signal into a register tree with a depth of $3$.
The automatic kernel scheduler of MaxCompiler ensured that every input to a given node in the design arrives at the correct cycle. If necessary, some streams were delayed, by the insertion of additional FIFOs or registers into the stream, until they synced up with the other streams. } 
Due to the finite amount of the available resources on the FPGA chip there is an upper bound in the number of \emph{Gate} blocks that can be implemented on the chip. 
In case the quantum circuit of our interest consists of more gates than it is implemented on the FPGA chip we make the stream $V$ to pass around the chain of the gate operations multiple times. 
The computational meta-data in the \emph{Gate} blocks (like the control and the target qubits of the transformation) are continuously updated according to the quantum gate sequence of the quantum circuit.
In order to buffer the transformed stream $V$, we incorporated the on-board memory installed on the FPGA card into the data-flow model as depicted in Fig. \ref{fig:building_blocks}.b).
The first \emph{Gate} block is getting the input $V_{\rm in}$ streamed from the on-board memory, and the last \emph{Gate} block is streaming the transformed matrix back into the on-board memory.
The buffering of the transformed unitary $V$ enables us to ensure continuous flow of data without stream conflicts in the \emph{Gate} blocks.
{The memory pool supporting the outlined buffering strategy is provided by $4\times16$ GB DDR4 DIMMs attached to the board of the Xilinx Alveo U250 FPGA card used in our work.
A single DIMM has normally a port width of $64$ bits and the basic burst size accessible over the bus is eight $64$-bit words, being the smallest addressable piece of memory in an FPGA implementation.
The bandwidth between the on-board memory and the FPGA fabric can be further increased by a factor of $4$ using all the DIMMS in parallel. 
The memory controllers (created by MaxCompiler and falling back to Xilinx IP blocks) creates monolithic memory blocks which are single address spaces. 
The memory addresses and commands to write or read the intermediate buffers are generated in advance on the chip according to the number of partial derivates to be evaluated (see bellow) and the size of the unitaries.}
Finally, the last \emph{Gate} block is extended with logic to evaluate the trace of the matrix $V$, and return the calculated value to the host CPU via the PCIe interface.
In order to evaluate the cost function (\ref{eq:frobenius}), we upload the $U^{\dagger}$ unitary describing the quantum program into the on-board memory in column-major order. 
The execution is started with streaming $U^{\dagger}$ towards the first \emph{Gate} block in the chain.
Due to the gate transformations the result after the final transformation would be $U^{\dagger}V$,  for which the trace is also calculated during the last clock cycles.

In our DFE implementation we could fit $108$ \emph{Gate} blocks on a single \emph{Super Logic Region} (SLR) of the FPGA chip.
(The $108$ \emph{Gate} blocks on the SLR are grouped into $6$ groups. 
The \emph{Gate} blocks within a group are executed synchronously, while FIFO buffered streams between the groups make it possible to run the groups asynchronously from each other easing the timing constraints along the long data paths.)
The Xilinx Alveo U250 FPGA card used in our numerical experiments contains $4$ of such SLRs, so in total a single DFE is capable to perform $432$ gate operations concurrently.
However, the connection between the SLRs is distant and uses only limited number of wires, implying the split of the implementation between SLRs and reduce the data transfer between these regions in order to meet the timing constraints even at higher clock frequencies. 
Hence, the DFE implementation is designed to simulate $4$ quantum circuits in parallel, each SLR doing an independent calculation. 
(During the optimization process the gradient components can be split between the SLRs, fitting to the concept of the developed DFE design.)

At this point we notice that the developed DFE implementation has a lower bound in the problem size than can be safely processed during the calculations.
When having a large degree of concurrency implemented on the chip (i.e. many \emph{Gate} blocks are chained up) the smaller matrices do not have enough elements to fully pipeline the gate chain and the implementation becomes stalled during execution.
In order to push the lower bound as low as possible we use a strategy of staggered data processing by mixing the calculation of the cost function (\ref{eq:frobenius}) with its partial derivatives. 
If the evaluation of the cost function takes multiple iterations over the gate chain, after each iteration we buffer the transformed matrix in the on-board memory, and continue the execution with the calculation of the gradient components.
(The evaluation of the gradient component takes the same number of gate operations as the cost function, only a single gate is replaced by its differentiated operation.)
The gradient components are also processed by a single iteration over the gate chain and the transformed unitaries gets buffered on the on-board memory.
This way the staggered data stream can be made long enough to adequately pipeline the DFE implementation. 
The memory addressing is governed by a memory address generator block supplying the memory controller with the correct memory addresses supporting the staggered data-flow design.
According to our experiments the smallest decomposition problem that can be safely processed by our design consists of $5$-qubits.
For the decomposition of smaller matrices pure CPU implementations can be used.

\paragraph*{Computational performance of the DFE implementation:} 

From computational performance point of view, the execution time on the DFE to evaluate once the cost function (\ref{eq:frobenius}) with its gradients can be calculated by the formula 
\begin{equation}
T_{\rm DFE}=(4^n\cdot (N_p+1))/(4\cdot f_{\rm DFE})\cdot{\rm ceil}(N_G/N_{G, chain}) + t_0,
\end{equation}
with $N_G$ standing for the number of quantum gates in the circuit, 
$N_{G, chain}$ being the number of \emph{Gate} blocks implemented in one SLR, $N_p$ labeling the number of the free parameters and $f_{\rm DFE}$ being the clock frequency running the gate operations.
Since the unitary matrix of an $n$-qubit quantum program contains $4^n$ elements, the first term in the formula gives the number of data elements processed under frequency $f_{\rm DFE}$.
(The function $\rm ceil()$ returns the next highest integer value by rounding up the argument if necessary.) 
The second term, $t_0$, describes the overhead time integrating the PCIe communication, CPU side data preparation and the initial pipelining of the gate chain.
In our build the frequency of the \emph{Gate} blocks is $f_{\rm DFE}=350$ MHz, while the \emph{gate kernel generator} blocks, the on-board memory controllers and address generator blocks are operating with frequencies $150$, $260$ and $250$ MHz, respectively.
(The frequency of the slower blocks were chosen on experimental basis by getting the first build meeting the time constraints. During the execution the slower blocks are in stalled state in most of the time.)  
For problem sizes of $8-9$ qubits $t_0$ is less than $1\%$ of the execution time, while for smaller problems it might take even the $30-50\%$. 
Parallel processing of CPU data preparation might significantly reduce this cost, but there is no workaround to lift the PCIe communication overhead, setting a hard limit of $\sim1$ms on $t_0$.

In order to make balance between on-chip resource usage and computational accuracy we used $32$-bit fixed point arithmetic operations in the implementation. 
Since the input matrix $U$ (along with the gate operations) is unitary, the elements in the data stream will not exceed the value of unity. 
Thus, we reserved one integer bit in the number representation to support this edge case, another bit is needed for the sign bit, while the remaining $30$ bits can be used for fractional bits.
To optimize the usage of digital signal processing (DSP) units in multiplications we followed the pioneering results of Refs.~\cite{8464809,5272296} and \cite{1388200} to optimally split the input multiplicands into smaller bitwidth parts and also under-utilizing the bitwidth where least common multiples maximize Karatsuba-style multiplication reductions based upon the width of the utilized input bitports of the DSP units.
(The DSP48E2 units embedded inside our Xilinx Alveo U250 FPGA cards have $18\times25$ wide input ports to perform signed multiplications and $16\times16$ and $17\times17$ tilings allow only 3 DSPs for $32$-bit $\times$ $32$-bit signed multiplication.)  We further apply the Knuth formula for 3M+5A complex multiplication \cite{knuth97} (i.e. $3$ real multiplications and $5$ real additions) as opposed to the standard 4M+2A which does not suffer numeric stability issues with fixed point and is no different if rounding $64$-bit sub-results to $32$-bit only after final summation.  This reduced the standard $16$ DSP solution down to only $9$ DSPs.
During the execution each \emph{gate block} performs $18$ arithmetic operations with $32$-bit fixed point numbers during a single clock cycle to apply the transformation shown in Fig. \ref{fig:building_blocks} in terms of two complex multiplications and a single complex addition.
Having $108$ \emph{gate blocks} on each SLR of the FPGA chip, the overall computational performance of the DFE is $2.72\times10^{12}$ $32$-bit fixed point operations per second (calculated with $f_{\rm DFE}=350$ MHz gate operation frequency), excluding all integer arithmetic-logical operations from the count.
Compared to CPU-like architectures this performance can be considered to be equivalent to $2.72$ TFLOPS.
Regarding the data transfer between the CPU host and the FPGA card we should notice that by having a separate gate chain on each of the SLRs, we need to provide each SLR with gate sequence data in parallel. 
In order to lift the congestion associated with the wiring of the data streams, we split each gate operation into four pockets up-streamed onto the chip sequentially. 
Thus, a valid gate operation is constructed during four clock cycles of the \emph{gate kernel generator} block.
In this way, the resources needed to stream up the gate sequences to each of the SLRs are reduced by a factor of four, while the \emph{gate kernel generator} block can still keep up with providing the $\emph{Gate}$ blocks with valid data without stalling.

Regarding the available resources of the chip, the usage of different hardware elements is correlated, since the individual blocks utilize various kind of hardware resources. The amount of the resources that can be utilized for the final implementation is limited by the growing on-chip congestion preventing to meet timing constraints for signal processing.
Our final FPGA build took $66.63\%$ of the DSP blocks available on the chip. 
The most intensively utilized components in our implementation are the BlockBRAM and URAM on-chip memory blocks with $83.31\%$ and $84.14\%$ usage, respectively.
The usage of other logical elements, like the look-up-tables (LUTs) and flip-flops (FFs) was $46.65\%$ and $43.63\%$, respectively.
Figure \ref{fig:building_blocks}.c) shows the area coverage on one of the SLR of the Xilinx Alveo U250 FPGA chip, the different colors indicate the individual blocks of the implementation. 
The memory controller and memory address generator are located in the center, and the $6$ \emph{Gate} block groups are placed around the central region. 

According to our numerical experiments, the DFE implementation on a single FPGA chip can evaluate the cost function (\ref{eq:frobenius}) and its gradient components by about $(6-13)\times$ faster than our CPU implementation on a 
computing server equipped with $32$-Core AMD EPYC 7542 Processor (providing $64$ threads with multi-threading) and with $128$GB of memory.
The computational speedup is more pronounced when the processed unitary is larger, or the quantum circuit is deeper. 
In such cases $t_0$ becomes less significant and the computations can be scaled up over multiple FPGA cards. 
In our numerical experiments summarized in Table.~1. we used $2$ FPGAs for the decomposition of $6$-qubit unitaries, $3$ FPGAs for $7$ and $8$ qubit unitaries (the $3$ FPGA cards were installed in a single host server), while for the decomposition of the $9$-qubit quantum program we have used two host servers equipped with $3$-$3$ FPGAs and connected with $100$ Gb/s network. 
During the execution we monitored $\sim30$ kB/s network traffic via MPI communication protocols by gathering the calculated gradient components.
In each case we observed ideal scaling of the execution time over the number of the utilized FPGA cards, up to $78\times$ speedup.

\section{Results}

Multi-qubit unitary operations associated with quantum programs can be decomposed in terms of single-qubit rotations and two-qunit controlled not ($CX$) gates\cite{PhysRevA.52.3457} (or other two-qubit gates supported by the actual physical realization of the quantum processor).
Since the number of two-qubit gates determines the upper bound of single-qubit gates in the circuit 
(for which none of the single-qubit gates can be merged into another one, while preserving the effect of the original circuit), the $CX$ gate count is one of the widely used metrics to benchmark quantum gate synthesis.
The $CX$ gate count accompanied with the depth of a circuit (i.e. the longest path in the circuit between the data input and the output while each gate counts as a unit) is an adequate way of characterizing the complexity of a circuit.

While deterministic synthesis tools \cite{qiskit_org,Sivarajah_2020,cirq} can be used to decompose larger unitaries on relatively short timescale (the synthesis of $7-9$ qubit unitaries took several minutes in our experiments), the produced circuit is in general too deep to perform successful experiments on NISQ quantum processors.
An inevitable ingredient to scale up quantum computations in the NISQ era relies on the ability of high-level circuit compression, holding the optimization based compilers\cite{qfast_github,qsearch_github,SQUANDER_github,cpflow_github} to high expectations.
However, currently the execution time of these compilers scales exponentially with the number of the involved qubits, hindering their large-scale utilization.
(Exception is made for variational quantum programs composed from smaller building blocks for which a pre-compiled gate structure can be adopted for the instantaneous variational parameter on a relatively short timescale\cite{squander2}.)

Our main achievement in this work is to develop a DFE implementation \cite{QGDDFE_github} supporting up to $9$-qubit circuits to speed up the simulation of the quantum circuits acting on the unitaries to be compiled. 
The developed accelerator might play a crucial role in sub-block optimization of large scaled quantum programs \cite{9743148,https://doi.org/10.48550/arxiv.2206.07885,https://doi.org/10.48550/arxiv.2206.13645}, that might be especially useful for variational quantum algorithms.
{The size of the sub-blocks in the partitioned circuit becomes a central issue: being able to optimize larger building blocks would result in a bigger compression rate on the addressed quantum program. In our research, we aimed to extend the maximum size of addressable quantum programs that could be decomposed without the need for partitioning them into smaller segments. 
At $9$ qubits the optimization landscape began to exhibit barren plateaus \cite{McClean2018}, which hindered further scaling of unitary decomposition for larger problems. In principle, the barren plateau problem could be overcome with a novel optimization strategy, as a fast cost function evaluation in itself still does not provide the final solution. However, the study of such an approach is outside the scope of the current work. Our research rather places the development of an FPGA accelerator in its focus to efficiently evaluate the cost function and its gradients.}
Instead of the lingering work of programming low level VHSIC Hardware Description Language (VHDL) components we used the high-level DFE development framework developed by Maxeler Technologies. 
Due to the programming components provided by the Java based DFE development framework we could design the quantum circuit DFE implementation in terms of high-level building blocks, such as support for arithmetic operations with complex numbers and trigonometric functions, stream-holds, memory controllers, etc.

By developing a combined DFE implementation of an arbitrary single- and controlled two-qubit gate operation we could increase the computational speed of the synthesis process by a factor of $(6-13)\times$ compared to the execution performance measured on a $32$-Core AMD EPYC 7542 Processor. 
The actual performance gain depends on the number of the qubits and on the number of free parameters of the circuit to be optimized. 
By splitting the calculations over multiple FPGA cards we could achieve in total $78\times$ speedup using $6$ FPGAs.
In principle, the performance can be further scaled up by incorporating additional FPGA cards. 
(As far the $1$ms initialization overhead of the FPGA card is significantly smaller than the computational time, the scaling of the performance with the number of the FPGA card is close to ideal.)

Exploiting the computational speedup we could extend the \emph{adaptive circuit compression} method of Ref.~\cite{squander2} to synthesize even $9$ qubit quantum programs with outstanding compression rate compared to the \emph{transpile} function of the QISKIT package. 
{At this stage it is important to emphasize that our work does not introduce algorithmic enhancements to the numerical approach of Ref.~\cite{squander2}. Nevertheless, to increase the readability of the paper, here we provide a brief overview on the key components of the algorithm.}

\begin{figure}
     \centering
     \includegraphics[width=0.9\textwidth]{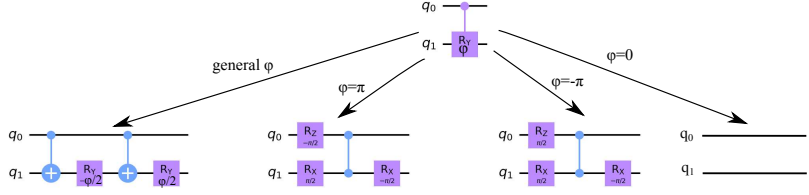} 
     \caption{Expansion of the controlled rotation $R_y(\varphi)$ into special gates according to the values of $\varphi$. For general values of the parameter $\varphi$ two $CX$ gates are needed to express the controlled rotation. For $\varphi=\pm\pi$ the effect of the controlled gate $R_y$ can be given by a single $CZ$ gate, while at $\varphi=0$ the gate is equal to identity. (The quantum circuit images were produced by the QISKIT package.)}
     \label{fig:CRY_expansion}
 \end{figure}

Following the reasoning of Ref.~\cite{squander2} the two-qubit gates used in the synthesis process are controlled rotation ($CR$) gates, tunable via a continuous parameter. 
The advantage of using such parametric two-qubit gates lies in their versatile ability to express quantum circuit elements, and the circuit compression approach would not be limited to local two-qubit gate cancellations.
At some specific parameter values the $CR$ gates can be considered as trivial, non-entangling gates, while at some other parameter values they can be mapped to special two-qubit gates such as $CX$ or controlled $Z$ ($CZ$) gate.
A special case of controlled rotations around the $y$ axis is shown in Fig.\ref{fig:CRY_expansion}.

Our synthesis results are summarized in Table \ref{table:results}., showing the circuit name and the number of the qubits taken from the online database \cite{ibm_mapping}, the number of $CX$ gates of the original decomposition and the obtained $CX$ count of the QISKIT and SQUANDER decompositions, respectively. 
{Unfortunately, optimization based quantum synthesis packages like QFAST and QSEARCH do not deliver sufficient computational performance to handle deep circuits such as reported in Table \ref{table:results}.
As for other widely used quantum computing software packages such as Tket, Cirq, PyQuil, Qulacs or Qibo, they incorporate efficient strategies for circuit optimization, but do not offer a functionality to decompose a quantum program described by a raw unitary matrix. 
Therefore we compared our results solely to the synthesized circuits produced by QISKIT.}

In our numerical experiments we assumed all-to-all connectivity between the qubits, however both packages could be executed for arbitrary connectivity topology as it was shown in Ref.~\cite{squander2}, preserving high compression rate provided by the optimization based methods.
Considering the depth metric, the circuits generated by SQUANDER are in average $19$ times shallower than the QISKIT generated circuits for the two $6$-qubit experiments addressed in our benchmark comparison.
The gap between the two implementations further increases with the number of qubits. 
For the six $7$-qubit unitaries and three $8$-qubit unitaries, the average compression of the circuits is $31$ and $324$-fold, respectively.
In our $9$-qubit experiment we achieved a $467$ times shallower circuit than with QISKIT.
Except a single example of the program \emph{rd53\_138}, our results also surpass the gate counts of the addressed circuits available at the online database \cite{ibm_mapping}.
(In order to provide fair comparison of the circuits we transformed the gates of the synthesized circuits into same gate set formed by general single qubit rotations and $CX$ gates. For the transformation we used the QISKIT package.)

\begin{table}
\begin{tabular}{||c | c | c | c | c | c | c | c | c | c | c||} 
 \hline
  Circuit name  & $n$ & \multicolumn{2}{c|}{IBM QX\cite{ibm_mapping}} & \multicolumn{2}{c|}{QISKIT\cite{qiskit_org}} & \multicolumn{3}{c|}{SQUANDER\cite{SQUANDER_github}} & comp. \\
      &  & $CX$ & $D$ & $CX$ & $D$ & $CX$ & $D$ & $f$ &  rate \\ [0.5ex] 
 \hline\hline
 4gt12-v0\_87 & $6$ & $112$ & $131$ & $625$ & $1146$ & $47$ & $73$ & $0.0028$  & $93.6\%$ \\ 
 4gt12-v0\_88 & $6$ &  $86$ & $108$ & $853$ & $1647$ & $44$ & $71$ & $0.0072$  & $95.7\%$ \\ 
 4mod5-bdd\_287 & $7$ &  $31$ & $41$ & $1037$ & $1825$ & $26$ & $41$ & $0.012$  & $97.8\%$ \\ 
 alu-bdd\_288 & $7$ &  $38$ & $48$ & $224$ & $408$ & $30$ & $35$ & $0.0038$  & $91.4\%$ \\ 
 C17\_204 & $7$ &  $205$ & $253$ & $2992$ & $5915$ & $102$ & $133$ & $0.0042$  & $97.8\%$  \\ 
ex2\_227 & $7$ &  $275$ & $355$ & $2852$ & $5554$ & $121$ & $161$ & $0.0128$  & $97.1\%$  \\ 
majority\_239 & $7$ & $267$ & $344$ & $4024$ & $7950$ & $141$ & $175$ & $0.0127$  & $97.8\%$  \\  
rd53\_131 & $7$ & $200$ & $261$ & $6538$ & $12320$ & $93$ & $119$ & $0.0129$ & $99.0\%$ \\
rd53\_135 & $8$ & $134$ & $159$  & $26126$ & $50436$ & $120$ & $147$ & $0.0195$ & $99.7\%$ \\
rd53\_138 & $8$ & $60$ & $56$  & $18567$ & $35172$ & $87$ & $117$ & $0.061$ & $99.7\%$ \\
cm82a\_208 & $8$ & $283$ & $337$ & $11246$ & $22284$ & $51$ & $67$ & $0.027$ & $99.7\%$ \\
con1\_216 & $9$ & $415$ & $508$ & $55822$ & $109798$ & $205$ & $229$ & $0.118$ & $99.8\%$ \\
 \hline
\end{tabular} 
\caption{$CX$ gate count and circuit depth ($D$) comparison of $n=6,\dots,9$ qubit circuits obtained by decomposing quantum programs taken from the circuits of the online database \cite{ibm_mapping}. The Fidelity $\overline{F}_F$ of the circuits is close to unity by an error of $10^{-4}$ (or less). The last column provides the compression rate in the circuit depth relative to the circuit synthesized by QISKIT.} \label{table:results}
\end{table}

{During our experiments, we observed that the decomposition of unitaries varied widely in terms of time, ranging from several hours for $6$-$7$ qubit problems to a week for the $9$-qubit scenario. 
(Without the implemented $6$-FPGA accelerator design the solution of the later problem would require approximately $13$ weeks on the three host servers, each of them equipped with dual socket $32$-core CPUs used in our experiments.) 

Our research indicates that as the number of qubits in the problem increases, the task of finding an appropriate decomposition becomes considerably more challenging. To achieve scalability for even larger problems, it would be required to develop a more efficient strategy for solving the optimization problem, rather than solely focusing on the acceleration to evaluate the cost function. However, the investigation and enhancement of the solver strategy lie outside the scope of the current manuscript.}

\section{Conclusions}  \label{sec:conclusion}

In this manuscript we reported on a hybrid CPU-FPGA quantum gate synthesis tool enabling one to extend the boundaries of optimization based quantum gate compilation from the earlier reported $6$ qubits up to $9$-qubit quantum programs.
The key ingredient to our solution is the FPGA based DFE implementation of a universal quantum computer simulator.
Our DFE implementation is fundamentally different from previous FPGA designs proposed to emulate quantum processors.
First of all, we are not pursuing the goal to scale up the number of qubits in our simulator like works \cite{Fujishima_2003,Goto_2007,Mahmud2019}.
Since the complexity of the optimization based synthesis process scales by $4^n$ with the number of qubits, we were after a reasonable trade-off between the implemented computational concurrency (in terms of on-chip multipliers used for $32$ bit multiplications) and the number of supported qubits (converted into on-chip memory usage) that is still reasonable from computational time point of view.
Secondly, for our calculations a universal implementation was needed supporting an arbitrary quantum circuit composed of single qubit rotations and conditional two-qubit gates, without the need to recompile the FPGA implementation when the gate structure changes. 
Our implementation is publicly available at a GitHub repository \cite{QGDDFE_github}.

To fulfill these requirements we designed a novel quantum computer simulator model developed by the high-level DFE development framework of Maxeler Technologies.
We did not intend to emulate the physical operation of quantum computers (like one-tick gate operations or simultaneous transformation of the state-vector amplitudes\cite{Khalid2021,Pilch2019}), but rather considered the design as a computational accelerator utilized in the synthesis process.
We used the developed DFE quantum computer simulator to decompose $6,7,8$ and $9$-qubit unitaries into a sequence of elementary gate operations accessible on NISQ devices, that was not shown before by optimization based gate synthesis tools.
Our numerical experiments using the SQUANDER package revealed outstanding circuit compression rate compared to the circuits generated by QISKIT, while keeping the fidelity of the synthesized circuits close to unity.
(The synthesized circuits generated by the SQUANDER package are accessible from the GitHub repository \cite{SQUANDER_github}.)
While $9$ qubits are still too few to demonstrate quantum advantage over classical devices, we think that larger circuits might be partitioned into few-qubit blocks that can be individually optimized by our approach \cite{9743148,https://doi.org/10.48550/arxiv.2206.07885,https://doi.org/10.48550/arxiv.2206.13645}.
We leave the study of this route for a future work.
As it was shown by recent benchmarks \cite{smith2021leap,squander2}, optimization based quantum compilers can also be used to synthesize highly optimized quantum circuit for topologies with limited connectivity as well, while the resulting circuit can be mapped to the gate types supported by the underlying hardware by simple transformations.
We believe that our contribution to the field is crucial to design shallow enough quantum circuits for NISQ devices to overcome the barrier of the error accumulation and demonstrate quantum mechanical problem solving at larger number of qubits. 
{Due to the high-level Python interface of SQUANDER it is possible to integrate the developed FPGA accelerator into other quantum packages as well.}

\section*{Acknowledgments}

We thank Masahiro Fujita and Vitali Averbukh for their valuable comments on the early version of the manuscript.
This research was supported by the Ministry of Culture and Innovation and the National Research, Development and Innovation Office within the Quantum Information National Laboratory of Hungary (Grant No. 2022-2.1.1-NL-2022-00004), by the ÚNKP-22-5 New National Excellence Program of the Ministry for Culture and Innovation from the source of the National Research, Development and Innovation Fund, and by the Hungarian Scientific Research Fund (OTKA) Grants No. K134437 and FK135220.
RP. acknowledge support from the Hungarian Academy of Sciences through the Bolyai J\'anos Stipendium (BO/00571/22/11) as well.
We acknowledge the computational resources provided by the Wigner Scientific Computational Laboratory (WSCLAB) (the former Wigner GPU Laboratory).

\appendix

\section{Brief description of the synthesis algorithm}

As we mentioned in the main text, our methodology of quantum gate synthesis is based on iterations of adaptive circuit compression during which the initially constructed over-parameterized quantum circuit is sequentially compressed by the removal of parametric two-qubit gates from the circuit. 
Following the reasoning of Ref.~\cite{squander2} the two-qubit gates used in the synthesis process are controlled rotation ($CR$) two-qubit gates, tunable via a continuous parameter. 
The advantage of using such parametric two-qubit gates lies in their versatile ability to express quantum circuit elements, and the circuit compression approach would not be limited to local two-qubit gate cancellations.
At some specific parameter values the $CR$ gates can be considered as trivial, non-entangling gates, while at some other parameter values they can be mapped to special two-qubit gates such as $CX$ or controlled $Z$ ($CZ$) gate.
A special case of controlled rotations around the $y$ axis is shown in Fig. 1 of  the main text.
This way it becomes possible to reformulate the structural combinatorial problem of placing the elementary two-qubit gates in a circuit into an optimization problem over continuous variables, which is the key concept to construct the initial quantum circuit processed by further compression iterations.  

In the very core of the optimization problem (either to construct the initial circuit or compress it) one need to evaluate a cost function describing the 'distance' of the synthesized $d\times d$ unitary $V$ from the original unitary $U$.
To this end Ref.~\cite{Khatri2019quantumassisted} introduced the Hilbert-Schmidt test
\begin{equation}
    C_{HST}(U,V) = 1 - \frac{1}{d^2}\left|\textrm{Tr}\left(V^{\dagger}U\right)\right|^2\,. \label{eq:hilbert-schmidt}
\end{equation}
The gate fidelity $\overline{F}(U,V)$, measuring the 'closeness' of two unitaries $U$ and $V$, is obtained
by averaging the state fidelities of output states (after a time evolution by $U$ and $V$ of the same initial state) over the Haar distribution  \cite{Khatri2019quantumassisted}.
\begin{equation}
    \overline{F}(U,V) = 1 - \frac{d}{d+1}C_{HST}(U,V)\;.
\end{equation}
Refs \cite{madden2021best,squander2,https://doi.org/10.48550/arxiv.2205.01121}, on the other hand, used a different, Frobenius norm based metric to quantify the distance between the two unitaries $U$ and $V$:
\begin{equation}
    f(U,V) = \frac{1}{2}\left\|V-U \right\|_F^2 = d - \textrm{Re}\left[{\textrm{Tr}}(VU^{\dagger})\right], \label{eq:frobenius_app}
\end{equation}
and defined a Frobenius based fidelity $\overline{F}_F(U,V)$ by
\begin{equation}
    \overline{F}_F(U,V) = 1 - \frac{d}{d+1} + \frac{1}{d(d+1)}\left( d - f(U,V)\right)^2 \label{eq:frobenius_fid}
\end{equation}
It can be shown that in general $\overline{F}_F(U,V)\leq \overline{F}(U,V)$ holds on\cite{madden2021best}.
In our numerical experiments we used Eq.~(\ref{eq:frobenius_app}) as the cost function during the optimization.
To find the minimum of the optimization problem we used the ADAM algorithm \cite{https://doi.org/10.48550/arxiv.1412.6980}, implying the need to calculate the gradient components of the cost function. 
Since the chosen cost function and all the quantum gates determining $V$ are linear operations, it is straightforward to derive the gradient of the cost function in terms of the free parameters:
\begin{equation}
    \frac{\partial f}{\partial x_i} = -\textrm{Re}\left[{\textrm{Tr}}\left(\frac{\partial V}{\partial x_i}U^{\dagger}\right)\right]\;.
\end{equation}
If the $x_i$ parameter is associated with the $k$-th quantum gate, the $\partial_i V$ gradient component of $V$ can be calculated by replacing the $k$-th gate with it's derivative with respect to $x_i$.
Since all the gates used in our approach contains trigonometric functions of the free parameters, the partial derivatives of the quantum gates can be obtained by the parameter shift rule
and by replacing the matrix elements independent of $x_i$ with zero.
The outlined method can be applied for both the single- and controlled two-qubit gates used in the quantum circuit.

To increase the success rate of the optimization process we developed several heuristic best-practices incorporated in the SQUANDER package.
First of all, to counterweight the potential of multi-start optimization shown in Refs \cite{smith2021leap,https://doi.org/10.48550/arxiv.2205.01121} we rather apply random shift to a portion of the parameters when the cost function seems to get caught in a local minimum or on a shallow plateau \cite{McClean2018,https://doi.org/10.48550/arxiv.2210.09191} and performing only a single instance of the optimization process.
Secondly, on occasional basis we increase the number of decomposing layers in the circuit (by adding new layers tuned to identity gates) and continue the optimization from the minimum obtained before the circuit was expanded.
This step can be combined with circuit compression iterations in order to increase the computational performance.
Finally, we should mention that Refs. \cite{https://doi.org/10.48550/arxiv.2205.04025} and \cite{https://doi.org/10.48550/arxiv.2210.09191} proposed modifications to the target cost function in order to increase the computational performance of quantum compilers. 
Their methods were shown to be efficient in several uses cases of shallow circuits. 
Unfortunately, in our numerical experiments involving deeper circuits we did not experience advantage in using these modifications.

\section{Numerical precision of quantum computer simulator implemented on DFEs}

In our implementation the DFE performs the calculations in $32$-bit fixed point number representation, providing less numerical accuracy compared to double precision floating point arithmetic typically used on CPU architectures.  
Though, our numerical experiments justified the choice of computational concurrency over the suppressed numerical precision:
the results summarized in Table 1. of the main text were not post-processed on CPUs to refine the parameters of the circuits, still achieving high gate fidelity of the synthesized circuits.
The accuracy of the cost function (1) of the main text starts to taper from double precision results only when getting close to the solution of the synthesis problem, otherwise the DFE holds up to accuracy of $6$ significant digits in the value of the calculated cost function.

\bibliographystyle{elsarticle-harv}
\bibliography{references}

\end{document}